\documentclass[twocolumn, aps, showpacs, showkeys, nofootinbib, floatfix, superscriptaddress]{revtex4}
\usepackage{amsfonts}

\usepackage{amssymb}
\usepackage{amsmath}
\usepackage{graphicx}
\begin{document}
\title{A more general interacting model of holographic dark energy}

\author{Fei Yu}
\email{yufei1980@mail.dlut.edu.cn}
\affiliation{School of Physics and Optoelectronic Technology, Dalian University of Technology, Dalian 116023, China}
\author{Jingfei Zhang}
\affiliation{Department of Physics, College of Sciences, Northeastern University, Shenyang 110004, China}
\author{Jianbo Lu}
\affiliation{School of Physics and Optoelectronic Technology, Dalian University of Technology, Dalian 116023, China}
\author{Wei Wang}
\affiliation{Physical Science and Technical College, Dalian University, Dalian 116622, China}
\author{Yuanxing Gui}
\affiliation{School of Physics and Optoelectronic Technology, Dalian University of Technology, Dalian 116023, China}

\begin{abstract}
So far, there have been no theories or observational data that deny the presence of interaction between dark energy and dark matter. We extend naturally the holographic dark energy (HDE) model, proposed by Granda and Oliveros, in which the dark energy density includes not only the square of the Hubble scale, but also the time derivative of the Hubble scale to the case with interaction and the analytic forms for the cosmic parameters are obtained under the specific boundary conditions. The various behaviors concerning the cosmic expansion depend on the introduced numerical parameters which are also constrained. The more general interacting model inherits the features of the previous ones of HDE, keeping the consistency of the theory.
\end{abstract}



\maketitle

\section{Introduction}
The observation of the type Ia Supernovae \cite{ref1} has shown that our universe is undergoing an era of accelerated expansion. Data from other observations such as Cosmic Microwave Background Radiation \cite{ref2} and SDSS \cite{ref3} also give support to the phenomenon. These data reveal that the present universe is dominated by 70\% exotic component, dubbed dark energy, which has negative pressure and is pushing the universe into accelerated expansion. Therefore, in order to explain this bizarre phenomenon, various models of dark energy have been put forward, ranging from the simplest one -- a cosmological constant -- to the scalar field theories of dark energy and modified gravity theories as well. The cosmological constant corresponds to the vacuum energy with constant energy and pressure, and an equation of state $w=-1$. However, it is confronted with two fundamental problems: the fine-tuning problem and the cosmic coincidence problem. To alleviate the drawbacks, various scalar field theories of dark energy emerge continually \cite{ref4}. On the other hand, the new geometric structures of space-time are also taken to realize the accelerated expansion of the universe \cite{ref5}.

Recently, another model inspired by the holographic principle has been put forward to explain the dark energy \cite{ref6,ref7}. The holographic principle, enlightened by the quantum properties of black holes, is one of the most important results in recent investigations of quantum gravity \cite{ref8,ref9}. It was first put forward by 't Hooft in the context of black hole physics \cite{ref10} and later extended to string theory by Susskind \cite{ref11}. According to the holographic principle, the entropy of a system scales not with its volume, but with its surface area \cite{ref12,ref13}. Equally the degrees of freedom of a piece of spatial region reside not in the bulk but only at the boundary of the region and the number of degrees of freedom per Planck area is not greater than unity. From the work of Cohen and collaborators \cite{ref14}, it is proposed that in quantum field theory a short distance cut-off is related to a long distance cut-off in virtue of the limit set by formation of a black hole. In other words, when $\rho_{de}$ is taken as the quantum zero-point energy density caused by a short distance cut-off, the total energy in the region of size $L$ is not more than the mass of a black hole of the same size, where the inequality $L^3\rho_{de} \leqslant LM_p^2$ emerges. Saturating it is equivalent to taking the largest $L$ allowed, thus
\begin{equation}\label{eq1}
\rho_{de}=3c^2M_p^2L^{-2},
\end{equation}
where $3c^2$ is a numerical constant introduced for convenience and $M_p$ is the reduced Planck mass, $M_p^2=(8\pi G)^{-1}$.

In the context of cosmology, the existence of an unknown vacuum energy is suggested and when $L$ is taken as the size of the current universe, for which the most common choice is the Hubble scale i.e., $L \sim H^{-1}$, the vacuum energy density is proportional to the square of the Hubble scale $\rho_{de} \propto H^2$, according to the holographic principle \cite{ref14,ref15}. (In this Letter we use terms like the vacuum energy and dark energy interchangeably.) Through the Friedmann equation
\begin{equation}\label{eq2}
3M_p^2 H^2=\rho_{de}+\rho_m,
\end{equation}
it turns out that the dark matter density $\rho_m$ has the same behavior as $\rho_{de}$ with the expression $\rho_m=3(1-c^2)M_p^2 H^2$. That is to say, the dark energy is also pressureless, namely, $p_{de}=w\rho_{de}=0$, meaning the equation of state parameter $w=0$. Obviously the case violates the condition of the cosmic accelerated expansion $w<-1/3$, although having solved the fine-tuning problem in principle \cite{ref16}. A better choice for the infrared cut-off $L$ is the particle horizon \cite{ref9,ref13}, which is defined as $R_{ph}=a\int_0^t\frac{dt'}{a}=a\int_0^a\frac{da'}{Ha'^2}$. But as is shown in Li's work \cite{ref6}, the equation of state of dark energy in this situation is larger than $-1/3$, out of the mechanism of cosmic acceleration likewise. Furthermore, Li takes the future event horizon, whose definition is $R_{eh}=a\int_t^\infty\frac{dt'}{a}=a\int_a^\infty\frac{da'}{Ha'^2}$, as the infrared cut-off instead, for the sake of ameliorating the holographic model. The aspect of amelioration is the satisfaction of acceleration and that $c=1$ makes the universe evolve into the de Sitter space-time while $c<1$ the phantom region \cite{ref17}. The concomitant defect is encounter of the causality problem pointed out by Cai \cite{ref18}. It is argued that according to the definition of the future event horizon, the history of dark energy depends on the future evolution of the scale factor $a(t)$, as violates causality. Moreover, for a spatially flat Friedmann-Robertson-Walker universe, only the cosmic expansion is accelerating can the future event horizon exist. So it is contrived that in order to interpret the cosmic accelerated expansion, the holographic dark energy model based on the future event horizon has presumed the acceleration. To avoid these problems originated by the introduced future event horizon, Cai also proposed a dark energy model, dubbed ``agegraphic dark energy'', characterized by the age of the universe, which was chosen as the length measure instead of the horizon distance of the universe. Correlative studies of agegraphic dark energy can be perused in \cite{ref19,ref20} and references therein. We give no unnecessary details here.

Recently, inspired by the holographic dark energy model, Gao et al. \cite{ref21} put forward the so-called Ricci dark energy (RDE) model. The idea is taking the average radius of Ricci scalar curvature $|R|^{-1/2}$ as the infrared cut-off. So from Eq. (\ref{eq1}) the dark energy density is proportional to the Ricci scalar curvature
\begin{equation}
\rho_{de}=3\alpha M_p^2\left(\dot{H}+2H^2\right)=-\frac{\alpha}{2}M_p^2 R,
\end{equation}
where $\alpha$ is a constant to be determined and a spatially flat universe is presumed. With some values of cosmological parameters which are consistent with current observations \cite{ref22,ref23}, the equation of state of Ricci dark energy can evolve across the cosmological constant boundary $w=-1$ \cite{ref24}. It means this model, which differs slightly for different values of cosmological parameters, can be classified as a quintom one \cite{ref25}. The Ricci dark energy model set up without the future event horizon is naturally free of the causality problem. In this case the dark energy is determined by the local Ricci scalar curvature rather than the event horizon of a global concept. Also, the fine-tuning problem is avoided because the dark energy is associated with the space-time scalar curvature, but not with Planck or other high energy physical scales and the coincidence problem is solved as well. In \cite{ref26}, Granda and Oliveros extended the Ricci dark energy to a more general form in which the energy density is
\begin{equation}\label{eq4}
\rho_{de}=3M_p^2(\alpha H^2+\beta\dot{H}),
\end{equation}
where $\alpha$ and $\beta$ are constants to be determined. So it can be realized that when $\alpha=2\beta$ the model reduces to Ricci dark energy. In the same way the model is phenomenologically viable, fitting with the current observational data, as well the causality and coincidence problems are solved. Correlative works of this kind are shown in \cite{ref27} from various perspectives of scalar field, spatial curvature, braneworld cosmology, observational data and so forth.

On the other hand, in the forementioned models, dark energy and dark matter evolve separately, keeping to different conservation equations of energy, with the standard evolution of dark matter $\rho_m \propto a^{-3}$. But so far, neither theories nor observational data have denied the interaction between them. For the sake of generality, the interaction term is naturally considered. In the context of dark energy, the interaction has been introduced to study some issues, e.g., raising accelerated expansion in an interacting dark energy model in which the Hubble scale is treated as the infrared cut-off \cite{ref28}, avoiding the big rip singularity \cite{ref29} and so forth \cite{ref30}.

Benefiting from investigations done already, we extend naturally the more general holographic Ricci dark energy model shown in \cite{ref26} to the form with interaction between the two major cosmic components, expecting to enrich the theoretical studies of series of holographic dark energy models. In the next section we expatiate upon the interacting model and obtain analytic expressions for the cosmic evolution. With the help of Suwa's constraints on the interacting Ricci dark energy (IRDE) model \cite{ref31}, we also constrain parameters in our model by analogism. The last section is for discussion.

\section{The interacting holographic dark energy model}
In the spatially homogeneous and isotropic universe the continuity equations of energy densities are given by
\begin{eqnarray}
\dot{\rho}_{de}+3H(1+w)\rho_{de} &=& -Q, \label{eq5} \\
\dot{\rho}_m+3H\rho_m &=& Q,
\end{eqnarray}
where $w$ is the equation of state of dark energy, i.e., $p_{de}=w\rho_{de}$. We take the interaction term of the form $Q=\Gamma\rho_{de}$, where $\Gamma=3b^2(1+r)H$ with the coupling constant $b^2$ and an introduced parameter $r=\rho_m/\rho_{de}$ as the ratio of two energy densities \cite{ref28}. Here the interaction is regarded as a decay process with an arbitrary decay rate $\Gamma$. Making use of the continuity equations we can get
\begin{equation}\label{eq7}
\dot{r}=3Hr\left(w+\frac{1+r}{r}\frac{\Gamma}{3H}\right).
\end{equation}
Likewise by the Friedmann equation (\ref{eq2}) the derivative of $H$ with respect to time is in the form
\begin{equation}\label{eq8}
\dot{H}=-\frac{3}{2}H^2\left(1+\frac{w}{1+r}\right).
\end{equation}
After the fractional energy densities have been introduced
\begin{equation}\label{eq9}
\Omega_m=\frac{\rho_m}{3M_p^2 H^2}, \ \ \
\Omega_{de}=\frac{\rho_{de}}{3M_p^2 H^2},
\end{equation}
the Friedmann equation has another expression
\begin{equation}
\Omega_m+\Omega_{de} = 1.
\end{equation}
Then substituting Eqs. (\ref{eq4}) and (\ref{eq8}) into Eq. (\ref{eq2}), we find the relationship between $w$ and $r$,
\begin{equation}\label{eq11}
w=\left(\frac{2\alpha}{3\beta}-1\right)(1+r)-\frac{2}{3\beta}.
\end{equation}
By the definition of $r$ and the Friedmann equation, $\Omega_{de}$ can be expressed in terms of $r$ as
\begin{equation}
\Omega_{de}=\frac{1}{1+r},
\end{equation}
and also the deceleration parameter
\begin{equation}
q=-1-\frac{\dot{H}}{H^2}=\frac{1}{2}+\frac{3w}{2(1+r)}.
\end{equation}

From the above expressions, it is noted that the main parameters, which can describe the evolution of the universe, appear as functions of the ratio $r$. So we turn to focus on the evolution of parameter $r$ and hope for some appropriate results. For a further step, when $w$ in Eq. (\ref{eq7}) is replaced by the expression (\ref{eq11}), it leads to a differential equation of $r$ with respect to $x=\ln{a}$,
\begin{equation}\label{eq14}
\frac{dr}{f(r)}=dx,
\end{equation}
where
\begin{eqnarray}
f(r) &=& \left(\frac{2\alpha}{\beta}-3+3b^2\right)r^2 \nonumber \\
& & +\left(\frac{2\alpha}{\beta}-3-\frac{2}{\beta}+6b^2\right)r+3b^2 \nonumber \\
&=& Cr^2+Br+A,
\end{eqnarray}
and $H=dx/dt$ is used in calculation. Integrate Eq. (\ref{eq14}),
\begin{equation}\label{eq16}
\int\frac{dr}{f(r)}=x+D,
\end{equation}
where $D$ is a constant of integration, which can be determined by the boundary condition $r_0=\rho_{m0}/\rho_{de0}=\Omega_{m0}/\Omega_{de0}$ and the subscript ``0'' denotes the current magnitude of the physical quantities. In order to solve Eq. (\ref{eq16}), we consider the discriminant of the quadratic polynomial $f(r)$, i.e., $\Delta=4AC-B^2$. But in this model there are three constants $\alpha$, $\beta$ and $b^2$ to be determined. We employ the same boundary conditions as authors of \cite{ref21,ref26} have done. At this rate, the number of parameters undetermined reduces to two.

Substituting boundary conditions $w_0=-1$ and $r_0=\frac{0.27}{0.73}$, which are consistent with current observations \cite{ref22,ref23}, into Eq. (\ref{eq11}), we obtain
\begin{equation}\label{eq17}
\alpha=\frac{2+3\beta r_0}{2(1+r_0)}.
\end{equation}
By the way, especially from the 7-year WMAP observations \cite{ref23}, the current data are consistent with a flat universe dominated by a cosmological constant, even when $w$ is dependent on time. Then $\beta$ and $b^2$ remain free and will be fixed by the behavior of the deceleration parameter. Combined with observations, the cases of $\Delta \geqslant 0$ are ruled out in the process of integration of Eq. (\ref{eq16}). Therefore, for $\Delta<0$ the evolution of the ratio of two energy densities with respect to $x$ is obtained as follows
\begin{equation}\label{eq18}
r(x)=\frac{\sqrt{-\Delta}\tanh\left[-\frac{\sqrt{-\Delta}}{2}(x+D)\right]-B}{2C},
\end{equation}
where the integration constant $D$ is
\begin{equation}\label{eq19}
D=-\frac{2}{\sqrt{-\Delta}}\tanh^{-1}\left[\frac{2Cr_0+B}{\sqrt{-\Delta}}\right].
\end{equation}

\begin{figure}
  \includegraphics[width=3 in]{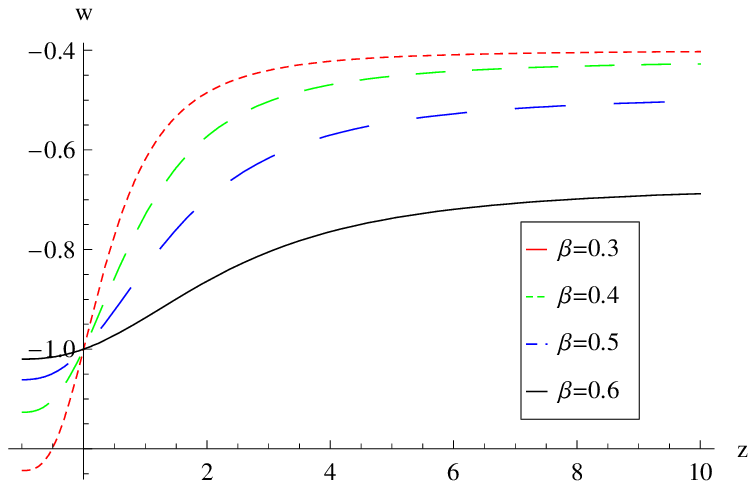}
  \includegraphics[width=3 in]{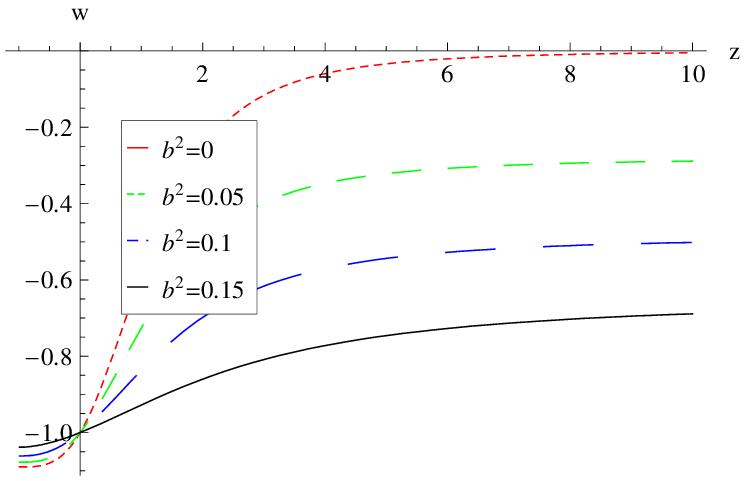}
  \caption{(Color online.) The evolution of the equation of state of dark energy $w$ with respect to the redshift $z$ for variable $\beta$ with the coupling constant $b^2=0.1$ and for variable $b^2$ with $\beta=0.5$. Herein $w_0=-1$, $\Omega_{m0}=0.27$ and $\Omega_{de0}=0.73$ have been used.}
\label{fig1}
\end{figure}

\begin{figure}
  \includegraphics[width=3 in]{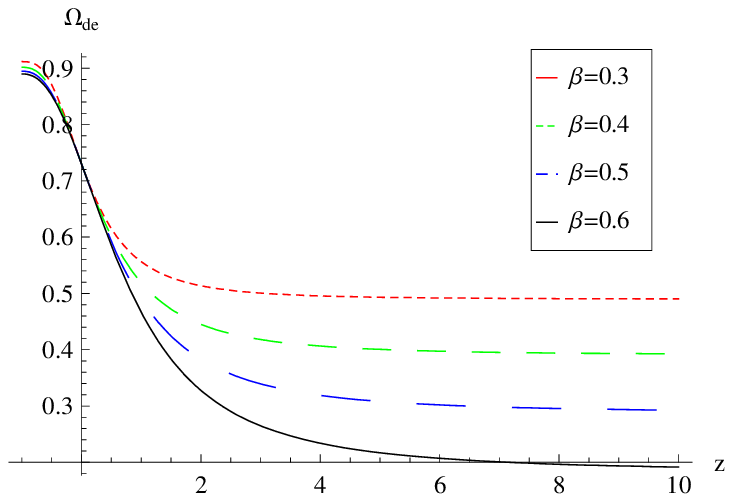}
  \includegraphics[width=3 in]{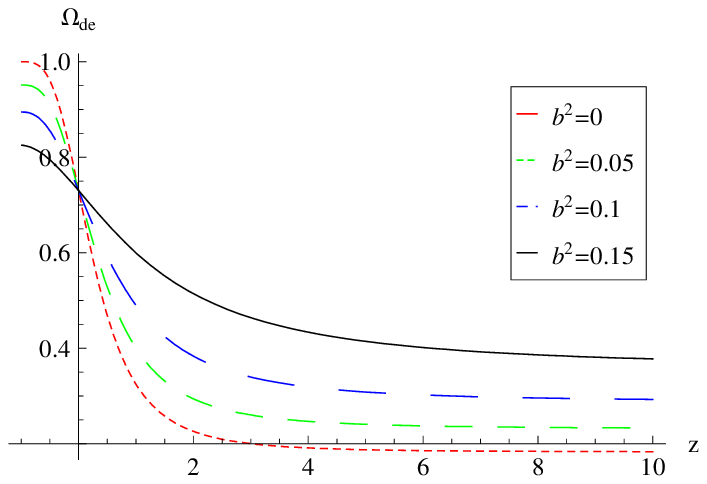}
  \caption{(Color online.) The evolution of the fractional energy density of dark energy $\Omega_{de}$ with respect to the redshift $z$ for variable $\beta$ with the coupling constant $b^2=0.1$ and for variable $b^2$ with $\beta=0.5$. Herein $w_0=-1$, $\Omega_{m0}=0.27$ and $\Omega_{de0}=0.73$ have been used.}
\label{fig2}
\end{figure}

\begin{figure}
  \includegraphics[width=3 in]{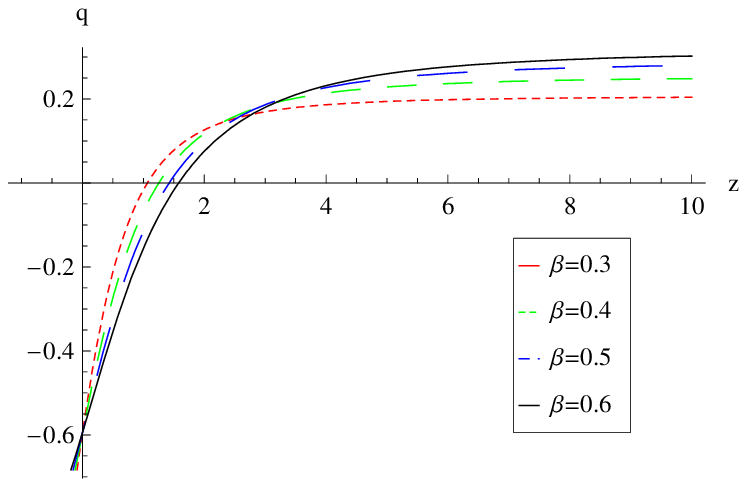}
  \includegraphics[width=3 in]{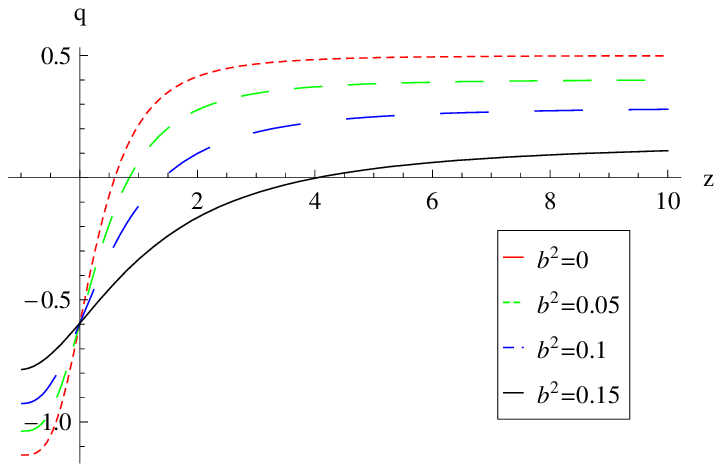}
  \caption{(Color online.) The evolution of the deceleration parameter $q$ with respect to the redshift $z$ for variable $\beta$ with the coupling constant $b^2=0.1$ and for variable $b^2$ with $\beta=0.5$. Herein $w_0=-1$, $\Omega_{m0}=0.27$ and $\Omega_{de0}=0.73$ have been used.}
\label{fig3}
\end{figure}

The evolutions of $w$, $\Omega_{de}$ and $q$ with respect to the redshift $z$ for variable $\beta$ with the coupling constant $b^2=0.1$ and for variable $b^2$ with $\beta=0.5$ are shown in Figs. \ref{fig1}, \ref{fig2} and \ref{fig3}, respectively. Under the boundary conditions used here, we note that $\alpha$ is independent of $b^2$ but only dependent of $\beta$, and increasing $\alpha$ for increasing $\beta$. When there is no interaction, i.e., $b^2=0$, the behaviors of the above parameters coincide with that in Ref. \cite{ref26}.

Most recently, Suwa and Nihei constrained the parameters in the IRDE model for the spatially flat universe by using the latest observational data from SNIa, combined with the CMB and BAO observations and the best-fit values given therein were $\Omega_{de0}=0.73\pm0.03$, $\alpha=0.45\pm0.03$ and $\gamma=0.15\pm0.03$ \cite{ref31}. The RDE is a quite typical model with $\alpha=2\beta$ in the more general one, therefore it is naturally interesting to take the IRDE model for a sample to constrain parameters in the general model.

To begin with, if we take $\alpha=2\beta$ into Eq. (\ref{eq11}) and use $\Omega_{de0}=0.73$, which is consistent with the boundary condition we use above, and $\beta=0.45\pm0.03$ for the corresponding coefficient of $\dot{H}$ in the definition of $\rho_{de}$, we will find that $\beta=0.46$ leads to $w_0\approx-0.993$ which is closest to $-1$ in the range for $\beta=0.45\pm0.03$. This coincidence nicely supports the boundary condition $w_0=-1$ we use above. Therefore, $\beta=0.46^{+0.02}_{-0.04}$ are taken for the best-fit values in the more general interacting holographic dark energy (IHDE) model.

Thus, by Eq. (\ref{eq17}) the best-fit values for $\alpha$ are $\alpha=0.9163^{+0.0081}_{-0.0162}$. However, it is worthwhile to declare that $\alpha$ is not necessarily equal to $2\beta$ any longer here in our model. Further, in light of the boundary conditions for the best-fit density ratio $r$ derived from literature \cite{ref31}, we can see about the coupling constant $b^2$ and the transition redshift $z_T$. In the past for $a\ll1$, $r\approx3.40$. If so, taking $\beta=0.46^{+0.02}_{-0.04}$, we get $b^2=0.001^{+0.024}_{-0.054}$ and $z_T=0.558^{+0.117}_{-0.170}$ as the best-fit values. For the case of $a\gg1$, namely in the future, $r\approx0.045$, then $b^2=0.046^{-0.001}_{+0.001}$ and $z_T=0.764^{+0.020}_{-0.046}$ are obtained. Note that there are two abnormal situations appearing. One is in the case of $a\gg1$, when $\beta$ increases to $0.48$, $b^2$ decreases by $0.001$ on the contrary while $\beta$ decreases to $0.42$, $b^2$ increases by $0.001$. It is opposite to the normal phenomenon, but we have gotten no clues yet for this. The other is in the case of $a\ll1$, $b^2<0$ appears numerically, which indicates that energy transfers from dark matter to dark energy in viewpoint of physics. Similar results have ever been obtained in Refs. \cite{ref32,ref33}. But the negative coupling constant will lead to consequences out of physics, like $\Omega_{de}$ will grow beyond $1$ in the far future. For this reason, we do not argue about the case of $b^2<0$, however, we favor the rational saying of elegancy from \cite{ref33} that {\it ``we let $b$ be totally free and let the observational data tell us the true story about the holographic dark energy, no matter whether the ultimate fate of the universe is ridiculous or not.''} So, though abnormality, we still retain the negative values of $b^2$ for a show. Also we have considered the non-interacting case, i.e., $b^2=0$, with $z_T=0.554^{+0.021}_{-0.041}$ gained. It is noted that the fitting values of $b^2$ and $z_T$ are consistent with the current observational data \cite{ref32,ref33,ref34}. Summarily, the best-fit values for parameters are displayed in Table \ref{table1}.

\begin{table}[h]
\caption{The best-fit parameters under different boundary conditions (BCs).}
\begin{tabular}{|c|c|c|c|}
     \hline\hline
     BCs & $r\approx3.40$ for $a\ll1$ & $r\approx0.045$ for $a\gg1$ & $b^2=0$ \\
     \hline
     $\Omega_{de0}$ & $0.73 \pm 0.03$ & $0.73 \pm 0.03$ & $0.73 \pm 0.03$ \\
     \hline
     $\alpha$ & $0.9163^{+0.0081}_{-0.0162}$ & $0.9163^{+0.0081}_{-0.0162}$ & $0.9163^{+0.0081}_{-0.0162}$ \\
     \hline
     $\beta$ & $0.46^{+0.02}_{-0.04}$ & $0.46^{+0.02}_{-0.04}$ & $0.46^{+0.02}_{-0.04}$ \\
     \hline
     $b^2$ & $0.001^{+0.024}_{-0.054}$ & $0.046^{-0.001}_{+0.001}$ & $0$ \\
     \hline
     $z_T$ & $0.558^{+0.117}_{-0.170}$ & $0.764^{+0.020}_{-0.046}$ & $0.554^{+0.021}_{-0.041}$ \\
     \hline\hline
\end{tabular}
\label{table1}
\end{table}

\begin{figure}
  \includegraphics[width=3 in]{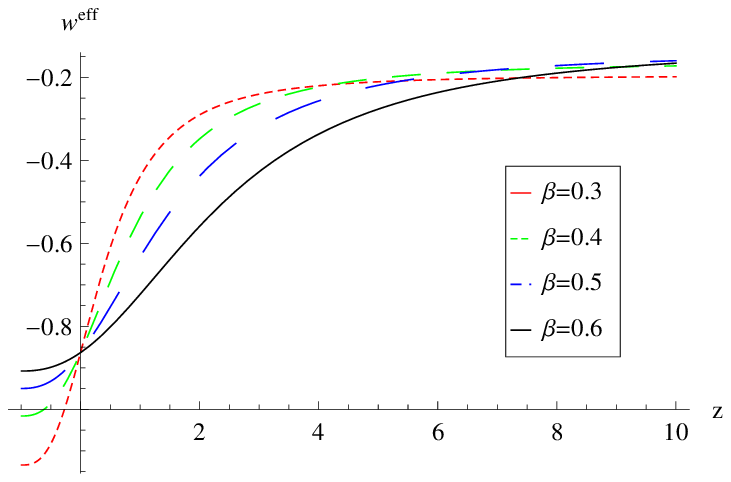}
  \includegraphics[width=3 in]{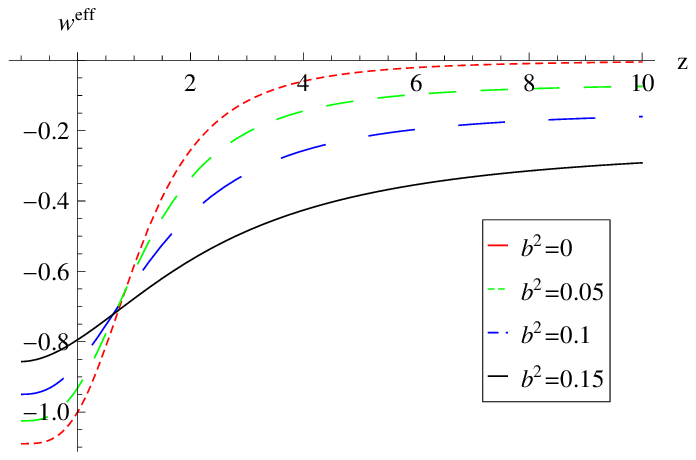}
  \caption{(Color online.) The evolution of the effective equation of state of dark energy $w^{\rm eff}$ with respect to the redshift $z$ for variable $\beta$ with the coupling constant $b^2=0.1$ and for variable $b^2$ with $\beta=0.5$. Herein $w_0=-1$, $\Omega_{m0}=0.27$ and $\Omega_{de0}=0.73$ have been used.}
\label{fig4}
\end{figure}

In our model, the boundary condition $w_0=-1$ is used, which makes the universe seem to behave like quintom, equal to cross the phantom dividing line, and to end with a big rip. But on the other hand, we consider the effective equation of state (EEoS) of dark energy to eliminate the unfavorable big rip singularity. Define the EEoS of dark energy as follows
\begin{equation}
w^{\rm eff}=w+\frac{\Gamma}{3H},
\end{equation}
then the continuity equation (\ref{eq5}) can be rewritten as
\begin{equation}
\dot{\rho}_{de}+3H(1+w^{\rm eff})\rho_{de}=0.
\end{equation}
The evolutions of $w^{\rm eff}$ with respect to the redshift $z$ for variable $\beta$ with the coupling constant $b^2=0.1$ and for variable $b^2$ with $\beta=0.5$ are shown in Fig. \ref{fig4}. The figures show clearly that for some values of $\beta$ and $b^2$, $w^{\rm eff}$ will remain greater than $-1$, featuring the quintessence-like behavior, which can avoid the future big rip singularity. Therefore, the interacting model is more favorable than the non-interacting one, although the interaction, even the evidence for it, between dark energy and dark matter is still not strong.

\section{Discussion}
Hereinafter, we review the consistency of formulae of the holographic dark energy density in various models. First of all, from the definition (\ref{eq9}) the dark energy density can be expressed generally $\rho_{de}=3M_p^2 \Omega_{de}H^2$. As is seen, if $\Omega_{de}$ is a constant, the energy density will correspond to the holographic model with the Hubble scale as the infrared cut-off, namely, $\rho_{de} \propto H^2$. It is understood easily. For $\rho_{de}=3c^2M_p^2 H^2$,
\begin{equation}
r\equiv\frac{\rho_m}{\rho_{de}}=\frac{1-c^2}{c^2}=\mbox{const.},
\end{equation}
which makes $\Omega_{de}=\frac{1}{1+r}=c^2$, so $\Omega_{de}$ must be a constant in this situation. Whereas, a variable $\Omega_{de}$, which is time dependent, will lead to other holographic dark energy models ever put forth. Concretely speaking, for the models with the particle horizon and the future event horizon, $L=R_h$ is presumed, where $R_h$ represents $R_{ph}$ or $R_{eh}$,
\begin{equation}
\Omega_{de}(t)=\frac{\rho_{de}}{3M_p^2 H^2}=\frac{3c^2M_p^2 R_h^{-2}}{3M_p^2 H^2}=\frac{c^2}{H^2 R_h^2},
\end{equation}
so $\Omega_{de}$ is time dependent. Also in the general model (\ref{eq4}), into which Eq. (\ref{eq8}) is  substituted for $\dot{H}$, the dark energy density reads
\begin{equation}\label{eq24}
\rho_{de}=3M_p^2\left[\alpha-\frac{3\beta}{2}\left(1+\frac{w}{1+r}\right)\right]H^2,
\end{equation}
where $r$ is in a state of dependence on time, and so is $w$. It means that terms in the square bracket of Eq. (\ref{eq24}) is a function of time, representing $\Omega_{de}(t)$.

Furthermore, we could see it from a different point of view. In \cite{ref28}, it is interestingly argued that for the Hubble scale case with interaction considered, if $c^2$ is time dependent, i.e., $c^2=c^2(t)$, the equation of state of dark energy will be
\begin{equation}
w=-\left(1+\frac{1}{r}\right)\left[\frac{\Gamma}{3H}+\frac{(c^2)^\cdot}{3Hc^2}\right].
\end{equation}
Thus both a varying $c^2$, which leads to a varying $r$, further a varying $\Omega_{de}$, and interaction can make $w$ more negative up to the cosmic accelerated expansion. In other words, the infrared cut-off may well remain $L=H^{-1}$, not necessarily changed. And the variation of $c^2(t)$ in $\rho_{de}=3c^2(t)M_p^2 H^2$ denotes the change of the degree of saturation of the holographic bound $\rho_{de} \leqslant M_p^2 L^{-2}$. It is equivalent to the above case $\rho_{de}=3M_p^2 \Omega_{de}(t)H^2$.

So far as it goes, these holographic dark energy models seem harmonious. However, as a matter of fact, although the infrared cut-off $L$ has been changed continually to solve problems such as fine-tuning, cosmic coincidence, accelerated expansion and causality step by step, the underlying physical mechanism of HDE remains vague yet. Therefore, we can but say that the series of HDE models are provided with consistency in sense of formalism and the internal relationship among them is claimed.

On the other hand, from analysis of curves above compared with that in \cite{ref35}, note that the increasing $b^2$ leads the range of $w$ and $\Omega_{de}$ to narrow down by reason of decay of the dark energy into dark matter, we have introduced up front. Also the larger $b^2$ is, the lower $w$ becomes in the early universe of high redshift, which means the behavior of dark energy is more different from that of dark matter, and the earlier the cosmic accelerated expansion occurs. But for smaller $b^2$, the cosmic expansion will gain larger acceleration in the far future. The conclusion here is coincident with that from Wang et al. \cite{ref35}.

In essence the dark energy problem should be an issue of quantum gravity, nevertheless there is no mature theory of quantum gravity with so many things unknown and uncertain at present. As regards the reason why the holographic dark energy density form containing a term $\dot{H}$ is motivated, Gao et al. bring forth their viewpoint from two perspectives of construction of a $K$-essence scalar field model and quantum fluctuation \cite{ref21}. Although the more general interacting holographic dark energy model acts phenomenologically viable, we sometimes avoid arguing round and round the subject phenomenologically and look forward to more penetrating insight into the holographic theory of dark energy.

\acknowledgments
This work is supported by the National Natural Science Foundation of China under Grant Nos. 10705041, 10975032 and 10947174.

\end{document}